\documentclass[aps,prb,twocolumn,showpacs,amsmath,amssymb,superscriptaddress]{revtex4}
\usepackage{graphicx}
\usepackage{amssymb}
\usepackage{natbib}

\begin{document}
\title{Phase diagram and neutron spin resonance of superconducting NaFe$_{1-x}$Cu$_x$As}

\author{Guotai Tan}
\affiliation{Department of Physics, Beijing Normal University, Beijing 100875, China}

\author{Yu Song}
\email{Yu.Song@rice.edu}
\affiliation{Department of Physics and Astronomy, Rice University, Houston, Texas 77005, USA}

\author{Rui Zhang}
\affiliation{Department of Physics and Astronomy, Rice University, Houston, Texas 77005, USA}

\author{Lifang Lin}
\affiliation{Department of Physics, Beijing Normal University, Beijing 100875, China}

\author{Zhuang Xu}
\affiliation{Department of Physics, Beijing Normal University, Beijing 100875, China}

\author{Long Tian}
\affiliation{Department of Physics, Beijing Normal University, Beijing 100875, China}

\author{Songxue Chi}
\affiliation{Quantum Condensed Matter Division, Oak Ridge National Laboratory, Oak Ridge, Tennessee 37831, USA}

\author{Barry Winn}
\affiliation{Quantum Condensed Matter Division, Oak Ridge National Laboratory, Oak Ridge, Tennessee 37831, USA}

\author{M. K. Graves-Brook}
\affiliation{Quantum Condensed Matter Division, Oak Ridge National Laboratory, Oak Ridge, Tennessee 37831, USA}

\author{Shiliang Li}
\affiliation{Beijing National Laboratory for Condensed Matter Physics, Institute of Physics, Chinese Academy of Sciences, Beijing 100190, China}
\affiliation{Collaborative Innovation Center of Quantum Matter, Beijing 100190, China}

\author{Pengcheng Dai}
\email{pdai@rice.edu}
\affiliation{Department of Physics and Astronomy, Rice University, Houston, Texas 77005, USA}
\affiliation{Department of Physics, Beijing Normal University, Beijing 100875, China}

\begin{abstract}
We use transport and neutron scattering to study the electronic phase diagram and spin excitations of
NaFe$_{1-x}$Cu$_x$As single crystals.  Similar to Co- and Ni-doped NaFeAs, a bulk superconducting phase appears near $x\approx2\%$ with the suppression of stripe-type magnetic order in NaFeAs. Upon further increasing Cu concentration the system becomes insulating, culminating in an antiferromagnetically ordered insulating phase near $x\approx 50\%$. Using transport measurements,
 we demonstrate that the resistivity in NaFe$_{1-x}$Cu$_x$As exhibits non-Fermi-liquid behavior near $x\approx1.8\%$.
Our inelastic neutron scattering experiments reveal a single neutron spin resonance mode exhibiting weak dispersion along $c$-axis in NaFe$_{0.98}$Cu$_{0.02}$As. The resonance is high in energy relative to the superconducting transition temperature $T_{\rm c}$ but weak in intensity, likely resulting from impurity effects. These results are similar to other iron pnictides superconductors despite the superconducting phase in NaFe$_{1-x}$Cu$_x$As is continuously connected to an antiferromagnetically ordered insulating phase near $x\approx 50\%$ with significant electronic correlations. Therefore, electron correlations is an important ingredient of superconductivity in NaFe$_{1-x}$Cu$_x$As and other iron pnictides. 
\end{abstract}


\maketitle


\section{Introduction}
Superconductivity in copper-oxides is obtained by electron- or hole-doping into their 
antiferromagnetically ordered insulating parent compounds with strong electron correlations \cite{PALee,keimer}. 
Superconductivity in iron pnictides, on the other hand, is derived from parent compounds that are bad metals \cite{kamihara,stewart,pdai}. 
Despite the metallic ground state of the parent compounds, strong electronic correlations are suggested to be present in iron 
pnictides \cite{QSi2008,HIshida2010,TMisawa2012}, especially in heavily hole-doped systems \cite{FHardy2013,LdeMedici2014,MNakajima2014}. In addition to replacing $A$ in $A$Fe$_{2}$As$_2$ ($A =$ Ca, Sr, Ba) with alkaline metals (K, Na, Rb, Cs) \cite{stewart,pdai}, hole-doping in iron pnictides can be achieved through Mn or Cr doping on Fe site although no superconductivity is induced in these cases \cite{AThaler2011,ASSefat2009}. Doping Cu on Fe site presents an intriguing case, whereas at low doping levels Cu is suggested to dope electrons \cite{SIdeta2013}, at higher concentrations Cu has a 3$d^{10}$ configuration and acts as a hole-dopant \cite{VKAnand2012,YJYan2013}. However, in Cu doped $A$Fe$_2$As$_2$, resistivity remains $\rho<1\text{m}\Omega\cdot\text{cm}$ throughout the whole phase diagram with no signs of strengthened electronic correlations \cite{YJYan2013}.  
As the As atoms within the same unit cell of $A$Fe$_{2-x}$Cu$_x$As$_2$ 
can form [As]$^{-3}\equiv$ [As-As]$^{-4}$/2 covalent bond with increasing 
Cu-doping and reducing As-As distance \cite{VKAnand2012,YJYan2013}, large Cu-doping therefore has little effect on the valence of Fe and 
$A$Cu$_{2}$As$_2$ becomes a $sp$-metal without electron correlations  
as predicted by band structure calculations \cite{DJSingh09}.

Since the crystal structure of NaFe$_{1-x}$Cu$_x$As 
does not allow the formation of As-As covalent bond [Fig. 1(a)] \cite{AFWang2013}, heavily doped NaFe$_{1-x}$Cu$_x$As is a candidate system to tune the strength of electronic correlations \cite{YSong}.
Recent transport \cite{AFWang2013}, scanning tunneling microscopy (STM) \cite{CYe2015}, angle-resolved photoemission spectroscopy \cite{CEMatt2016}, and optical conductivity \cite{ACharnukha} measurements demonstrated that with significant ($>10\%$) Cu doping, NaFe$_{1-x}$Cu$_x$As acquires an insulating ground state. Diffraction measurements revealed short-range Fe-Cu cation order and magnetic order develops as the system becomes insulating. With increasing $x$, the Fe-Cu cation order and magnetic order continuously increase in correlation lengths, and become long-range when $x\approx50\%$ \cite{YSong}. The antiferromagnetically ordered insulating state 
in NaFe$_{1-x}$Cu$_x$As is a result of significant electronic correlations \cite{YSong}. 
Compared to Cu-doped $A$Fe$_2$As$_2$, NaFe$_{1-x}$Cu$_x$As is more correlated because of its larger iron pnictogen height which 
results in a small magnetic excitation bandwidth \cite{ZPYin2011,CZhang2013}, hole-doping effect of Cu in NaFe$_{1-x}$Cu$_x$As is not counteracted by the formation of As-As covalent bonds and the local potential differences between Cu and Fe reduces the hopping between Fe sites \cite{YSong}. 
Since the antiferromagnetically ordered Mott insulating phase of NaFe$_{1-x}$Cu$_x$As at $x\approx 0.5$ is continuously connected to an albeit somewhat far away superconducting phase at $x\approx 0.02$ \cite{AFWang2013,YSong}, it would be important to elucidate 
whether the superconducting state in NaFe$_{1-x}$Cu$_x$As \cite{AFWang2013} is similar to other iron pnictide systems.

One hallmark of unconventional superconductivity in iron pnictides is the
appearance of a neutron spin resonance mode in the superconducting state at the antiferromagnetic (AF) ordering
wave vector of their parent compounds
\cite{christianson,pdai,scalapino,DSInosov2011}.  The energy of the resonance approximately scales with the superconducting transition temperature $T_{\rm c}$ or the superconducting gap $\Delta$ \cite{DSInosov2011,GYu}. The appearance of the resonance mode is typically accompanied by gapping of the normal state spectral weight below the resonance energy \cite{christianson,pdai,scalapino,DSInosov2011}. Both the resonance and the associated spin gap are typically interpreted as due to quasi-nested Fermi surfaces that results in a collective bound-state inside the superconducting gap \cite{scalapino,Eschrig,PJHirschfeld}. Within this picture, the resonance mode is evidence for unconventional superconductivity with sign-changing superconducting order parameters on different parts of the Fermi surface \cite{christianson}. The resonance in BaFe$_2$As$_2$-derived superconductors display significant dispersion along $c$-axis in the underdoped regime \cite{SChi09}, and becomes $L$-independent in the well-overdoped region \cite{CHLee2013}. In electron-doped NaFe$_{1-x}$Co$_x$As, two resonance modes are seen in underdoped compositions \cite{CZhang2013}, with the lower-energy-mode gradually losing spectral weight upon further doping before disappearing near optimal doping \cite{CZhang2016,CZhang2013b}. The energy of the single resonance mode in well-overdoped NaFe$_{1-x}$Co$_x$As does not scale with $T_{\rm c}$, likely due to multi-orbital physics and impurity effects \cite{CZhang2016}. Since superconducting domes are absent in Cu-doped $A$Fe$_2$As$_2$ \cite{NNi2010,YJYan2013}, 
it is unclear if superconductivity induced by Cu-doping has similar electronic and magnetic properties 
as their electron/hole-doped counterparts. As NaFe$_{1-x}$Cu$_x$As exhibits bulk superconductivity similar
to those of NaFe$_{1-x}$Co$_x$As \cite{AFWang2013}, superconducting NaFe$_{1-x}$Cu$_x$As offers a unique case to study the effect of the stronger impurity potential of Cu \cite{AFWang2013,MGKim2012} on the resonance mode.

In this work, we use transport and magnetic susceptibility measurements to characterize NaFe$_{1-x}$Cu$_x$As with $x\leq0.06$ and inelastic neutron scattering to study the neutron spin resonance in slightly overdoped NaFe$_{0.98}$Cu$_{0.02}$As. The temperature dependence of resistivity display an evolution from Fermi liquid behavior in NaFeAs, to non-Fermi-liquid behavior near $x=0.018$ and back to Fermi-liquid behavior near $x=0.032$, pointing to a funnel of quantum critical behavior near optimal doping. A single resonance mode that disperses weakly along $c$-axis is seen in NaFe$_{0.98}$Cu$_{0.02}$As, similar to slightly overdoped NaFe$_{1-x}$Co$_x$As \cite{CZhang2013b}. Similar to what was found in heavily overdoped NaFe$_{1-x}$Co$_x$As \cite{CZhang2016}, the resonance mode is at an higher energy than what is expected from scaling with $T_{\rm c}$, likely related to the stronger impurity potential of Cu compared to Co \cite{AFWang2013,MGKim2012}, despite the low Cu concentration of $2\%$. The presence of both a putative quantum critical point in the phase diagram and a neutron spin resonance in superconducting NaFe$_{1-x}$Cu$_x$As is similar to other iron pnictide superconductors. Since NaFe$_{1-x}$Cu$_x$As can be continuously tuned to an antiferromagnetically ordered insulating phase with significant electronic correlations near $x\approx0.5$ \cite{YSong}, these results suggest electronic correlations to be an essential ingredient of superconductivity in iron pnictides. 

\begin{figure}[t] 
\includegraphics[scale=.38]{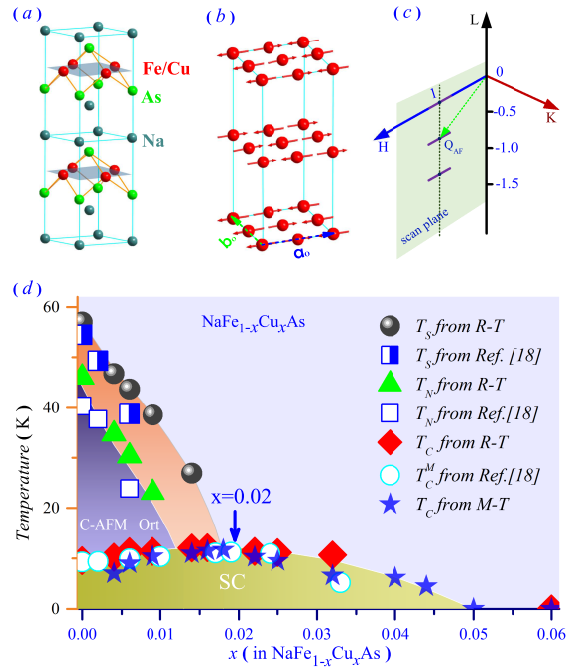}
\caption{(Color online)
(a) Crystal structure of NaFeAs, two tetragonal unit cells stacked along the $c$-axis are shown. The orthorhombic unit cell is twice the size of the tetragonal unit cell with $a_{\rm O}\approx\sqrt{2}a_{\rm T}$ and $b_{\rm O}\approx\sqrt{2}b_{\rm T}$ and rotated by 45$^{\circ}$ compared to the tetragonal unit cell. (b) The magnetic structure of NaFeAs, only Fe atoms are shown. The magnetic unit cell is twice the size of the orthorhombic unit cell along $c$-axis. (c) Schematic of $[H,0,L]$ scattering plane. (d) Phase diagram of NaFe$_{1-x}$Cu$_x$As obtained from magnetic susceptibility and resistivity measurements, results from previous work \cite{AFWang2013} are shown for comparison. The vertical arrow marks $x=0.02$, for which inelastic neutron scattering measurements were carried out.
}
\end{figure}

\begin{figure}[t] 
\includegraphics[scale=.45]{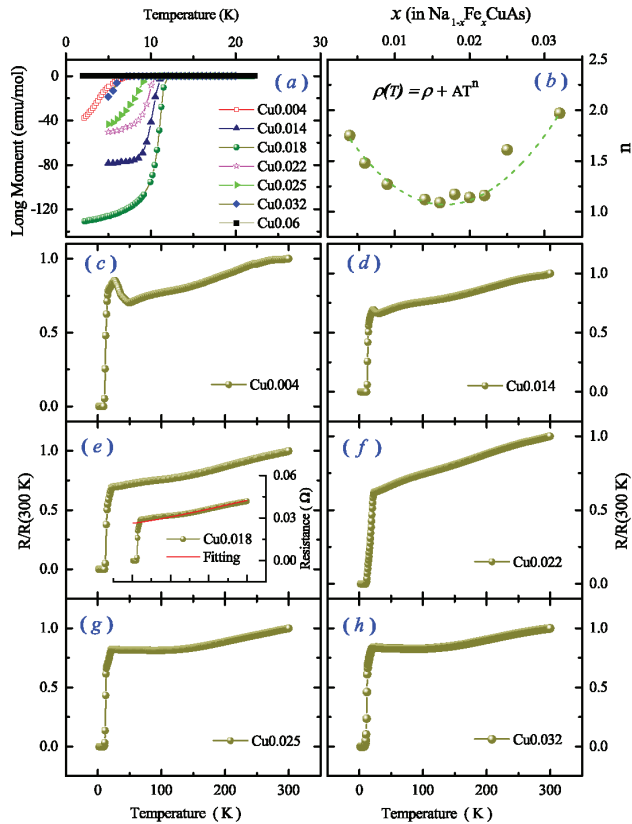}
\caption{(Color online)
(a) Magnetic susceptibility of NaFe$_{1-x}$Cu$_x$As measured upon warming after zero-field-cooling. (b) $n$ extracted from fitting resistivity of NaFe$_{1-x}$Cu$_x$As to the form $\rho=\rho_0+AT^n$, the solid line is a guide-to-the-eye. (c)-(h) respectively shows resistivity of NaFe$_{1-x}$Cu$_x$As with $x=0.004$, 0.014, 0.018, 0.022, 0.025 and 0.032 normalized to room temperature resistivity. The inset in (e) shows raw data for $x=0.018$ and a typical fit to the empirical form mentioned in the text.
}
\end{figure}

\section{Experimental results}

Single crystals of NaFe$_{1-x}$Cu$_x$As were prepared using the self-flux method method as 
described in previous work \cite{NSpyrison2012}. Inductively coupled plasma atomic emission spectroscopy previously revealed the actual Cu concentration to be similar to the nominal one for $x\lesssim20\%$ \cite{YSong}, we therefore quote nominal concentrations throughout this work, which is also consistent with previous report on NaFe$_{1-x}$Cu$_x$As in the superconducting region \cite{AFWang2013}. Resistivity and magnetic susceptibility measurements were carried out using commercial systems from Quantum Design. Resistivity is measured with the four-probe method, and magnetic susceptibility is measured upon warming with an in-plane magnetic field of 20 Oe after zero-field-cooling. Inelastic neutron scattering experiments were carried out using the HB-3 triple-axis spectrometer at the High Flux Isotope Reactor (HFIR) and the Hybrid Spectrometer (HYSPEC) at the Spallation Neutron Source, both at Oak Ridge National Laboratory. The experiment at HB-3 used a pyrolitic graphite monochromtor, analyzer and filter after the sample, the collimation used is 48$^\prime$-40$^\prime$
-sample-40$^\prime$-120$^\prime$ and the final neutron energy is fixed to $E_{\rm f}=14.7$ meV. HYSPEC is a time-of-flight chopper spectrometer with a movable strip-shaped detector bank that has much fewer pixels along the vertical direction compared to the horizontal direction. Detected neutron counts from the middle third of the pixels along the vertical direction are binned making it function like a triple-axis spectrometer with a position-sensitive-detector, and by rotating the sample and the detector bank, maps of the scattering plane can be obtained. Fixed incident energy $E_{\rm i}=15$ meV is used for the experiment on HYSPEC. Momentum transfer ${\bf Q}=(Q_x,Q_y,Q_z)$ is presented in reciprocal lattice units (r.l.u.) as $(H,K,L)$, with $H=Q_x a/2\pi$, $K=Q_y b/2\pi$ and $L=Q_z c/2\pi$. We adopt the chemical unit cell for the orthorhombic phase of NaFe$_{1-x}$Cu$_x$As, in this notation $a\approx b\approx5.56$ and $c\approx7.05$ for NaFeAs \cite{sli}. The orthorhombic unit cell is twice the volume of the tetragonal unit cell [Fig. 1(a)], in this notation magnetic Bragg peaks are seen at ${\bf Q}=(1,0,L)$ with $L=0.5,1.5,2.5\ldots$, corresponding to half of the magnetic unit cell along $c$-axis [Fig. 1(b)]. 8 high-quality single crystals with a total mass of 6.04 grams were co-aligned in the $[H,0,L]$ scattering plane [Fig. 1(c)].     

\begin{figure}[t] 
\includegraphics[scale=.5]{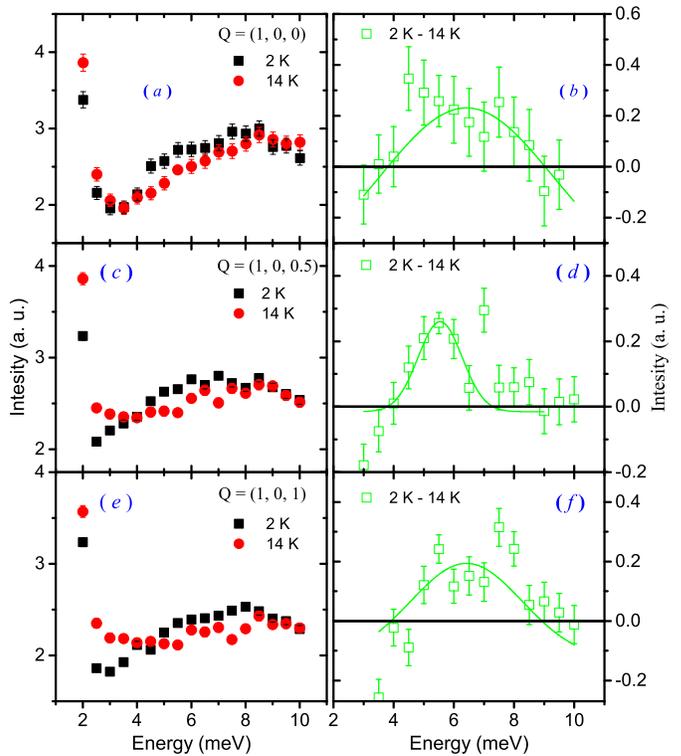}
\caption{(Color online) Constant-${\bf Q}$ scans at ${\bf Q}=(1,0,L)$ for (a) $L=0$, (c) $L=0.5$ and (e) $L=1$. The difference of magnetic intensity in the superconducting state and the normal state for (b) $L=0$, (d) $L=0.5$ and (f) $L=1$. The solid lines are fits to Gaussian peaks, which provide a rough estimate of the peak center of the resonance mode. The fits in (b) and (f) are constrained to have the same center.
}
\end{figure}

Figure 2 summarizes magnetic susceptibility and resistivity measurements. As can be seen in Fig. 2(a) optimal superconductivity with $T_{\rm c}\approx11.8$ K is obtained near $x=0.018$ in agreement with previous results \cite{AFWang2013}. Normalized resistivity in NaFe$_{1-x}$Cu$_x$As with $x=0.004$, 0.014, 0.018, 0.022, 0.025 and 0.032 are shown in Fig. 2(c) through Fig. 2(h). For underdoped samples ($x=0.004$ and 0.014), resistivity exhibits clear kinks at $T_{\rm s}$ and $T_{\rm N}$ before vanishing in the superconducting state below $T_{\rm c}$.  Resistivity data in the tetragonal paramagnetic metallic state are fit to the empirical form $\rho=\rho_0+AT^n$ for all measured samples [inset of Fig. 2(e)], similar to previous work on 
NaFe$_{1-x}$Co$_x$As \cite{GTan2016,AFWang2013b}. A Fermi-liquid corresponds to $n=2$, whereas linear resistivity ($n=1$) is often observed near a quantum critical point \cite{SKasahara2010,JGAnalytis2014}. Doping dependence of $n$ for NaFe$_{1-x}$Cu$_x$As is summarized in Fig. 2(b), revealing clear evolution from $n\approx2$ in NaFeAs to $n\approx1$ for $x=0.018$, and back to $n\approx2$ for $x=0.032$. Similar behavior has been observed in other iron pnictide systems \cite{SKasahara2010,JGAnalytis2014,GTan2016,AFWang2013b,HLuo2012}. Based on magnetic susceptibility and resistivity measurements, we construct the phase diagram of NaFe$_{1-x}$Cu$_x$As near the superconducting dome as shown in Fig. 1(d), the obtained phase diagram is consistent with previous results \cite{AFWang2013}.

\begin{figure}[t] 
\includegraphics[scale=.5]{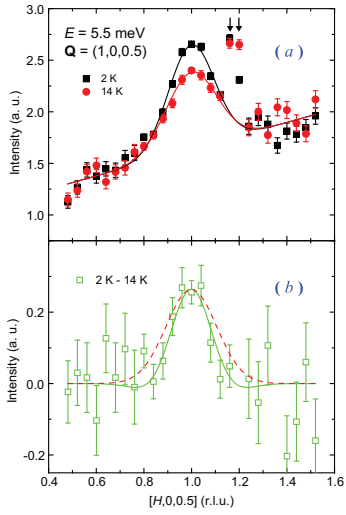}
\caption{(Color online) (a) $H$-scans at $E=5.5$ meV centered at ${\bf Q}=(1,0,0.5)$. Solid lines are fits to Gaussian peaks with the linear background restrained to be the same for 2 K and 14 K. The two positions marked by arrows are contaminated by spurious scattering and are not used in the fits. (b) The difference between 2 K and 14 scans in (a), the solid line is the difference between the fits at 2 K and 14 K. The red dashed line is a Gaussian peak with width of the data at 14 K, shown for comparison. Fitting the difference to a Gaussian peak [not shown]  results in a much narrower peak [FWHM = 0.17(3) (r.l.u.)] compared to the 14 K data [FWHM = 0.25(1) (r.l.u.)].}
\end{figure}

Constant-${\bf Q}$ scans at ${\bf Q}=(1,0,L)$ for slightly overdoped NaFe$_{0.98}$Cu$_{0.02}$As ($T_{\rm c}=11.8$ K and no static magnetic signal is observed) measured using HB-3 are summarized in Figure 3. Scans above ($T=14$ K) and below ($T=2$ K) $T_{\rm c}$ for $L=0$, 0.5 and 1 are shown in Fig. 3(a), (c) and (e), respectively. The corresponding 2 K data after subtracting 14 K data are similarly shown in Fig. 3(b), (d) and (f). A clear resonance mode that displays a weak $c$-axis dispersion accompanied by a spin gap at lower energies can be clearly seen. For $L=0.5$ [Fig. 3(d)], corresponding to the magnetic zone center in magnetically ordered NaFeAs, the resonance is centered at around $E\approx5.5$ meV. Similarly for $L=0$ and 1  [Fig. 3(b), (f)], corresponding to the magnetic zone boundary along $c$-axis in magnetically ordered  NaFeAs, the resonance is centered at $E\approx6.5$ meV. The resonance mode at $L=0.5$ also appears to be sharper than for $L=0$ and 1, this behavior is different from slightly overdoped NaFe$_{0.955}$Co$_{0.045}$As in which width of the resonance mode does not depend on $L$ \cite{CZhang2013b}.  

\begin{figure}[t] 
\includegraphics[scale=.4]{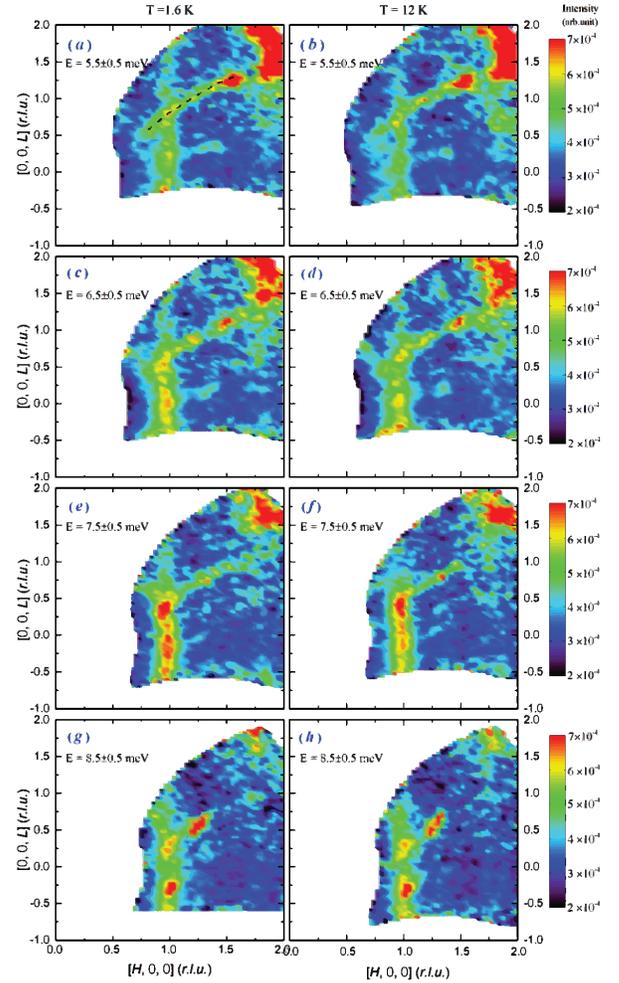}
\caption{(Color online) Constant-energy maps in $[H,0,L]$ scattering plane at $T=1.6$ K for (a) $E=5.5\pm0.5$ meV, (c) $E=6.5\pm0.5$ meV, (e) $E=7.5\pm0.5$ meV and (g) $E=8.5\pm0.5$ meV. The corresponding results at $T=12$ K are shown in panels (b), (d), (f) and (h). The streak of signal marked by the dashed line in (a) is spurious, as are similar signals in other panels. 
}
\end{figure}

$H$-scans at the peak of the resonance mode for $L=0.5$ ($E=5.5$ meV) is shown in Fig. 4. At both 2 K and 14 K, a clear peak is observed, with the peak at 2 K only slightly stronger than the one at 14 K [$\approx 23\%$ stronger from area of fits in Fig. 4(a)], meaning the resonance is weak compared to the normal state excitations in contrast to slightly overdoped NaFe$_{1-x}$Co$_x$As \cite{CZhang2013b} in which the resonance mode dominates the magnetic excitations in the superconducting state. Given the normal state spin fluctuations have indistinguishable intensities in Co- and Cu-doped BaFe$_2$As$_2$ \cite{MGKim2012}, it is reasonable to assume the normal state intensities in Co- and Cu-doped NaFeAs are also similar. Therefore, compared to NaFe$_{0.955}$Co$_{0.045}$As \cite{CZhang2013b}, the resonance mode in NaFe$_{0.98}$Cu$_{0.02}$As is also quantitatively much weaker. The width of the magnetic peak for $E = 5.5$ meV at 2 K [FWHM = 0.228(8) (r.l.u.)] and 14 K [FWHM = 0.25(1) (r.l.u.)] are similar with the peak in the superconducting state slightly narrower, similar to slightly overdoped NaFe$_{1-x}$Co$_x$As \cite{CZhang2013b}. The change in peak width can be seen more clearly by examining the difference of 2 K and 14 K data [Fig. 4(b)], which is significantly narrower than the 14 K data itself [Gaussian peak in red dashed line], with FWHM = 0.17(3) (r.l.u.). This suggests that the effect of superconductivity is not only to enhance intensity at the energy of the resonance mode, but also to increase the correlation length of magnetic excitations. The resonance being more well-defined in momentum space compared the normal state excitations can result from reduction of damping due to opening of the superconducting gap, or the resonance mode intrinsically having a longer correlation length. A weak neutron spin resonance mode is seen in Ba(Fe$_{1-x}$Ru$_x$)$_2$As$_2$ near optimal doping and ascribed to weakened electronic correlations \cite{JZhao2013}. In comparison, the weak resonance in NaFe$_{0.98}$Cu$_{0.02}$As is likely due to impurity effects given electronic correlations in superconducting NaFe$_{1-x}$Co$_x$As and NaFe$_{1-x}$Cu$_x$As should be similar.

\begin{figure}[t] 
\includegraphics[scale=1]{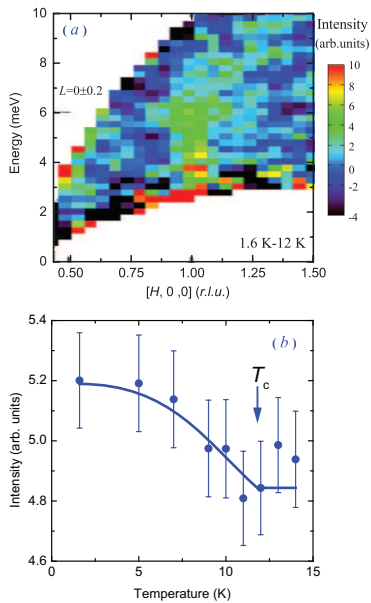}
\caption{(Color online)
(a) $H$-energy map of 1.6 K data subtract by 12 K data obtained by binning data with $-0.2\leq L\leq0.2$. (b) Temperature dependence of the resonance mode, obtained by binning data with $4.5\leq E\leq6.5$ meV, $0.87\leq H\leq1.13$ and $-0.2\leq L\leq0.2$. The arrow marks $T_{\rm c}$, and the solid line is a guide-to-the-eye.
}
\end{figure}

Our results are further substantiated by data obtained using HYSPEC in Figure 5 and 6. Constant-energy maps of the $[H,0,L]$ scattering plane are shown in Figure 5 at $T=1.6$ K and 12 K, as can be clearly seen magnetic excitations form rods centered at $H=1$ with little $L$-dependence in both the normal and the superconducting state. In magnetically ordered NaFeAs, $c$-axis polarized spin waves exhibit a spin gap of $E_{\rm g}\approx 4.5$ meV at ${\bf Q}=(1,0,0.5)$ and $E_{\rm g}\approx7$ meV at ${\bf Q}=(1,0,1)$ \cite{YSong2013}. Upon Co-doping into the slightly overdoped regime, magnetic excitations become $L$-independent \cite{CZhang2013b}. Therefore, the reduction of effective $c$-axis magnetic coupling by doping Cu doping into NaFeAs is similar to doping Co, such behavior is also seen in Ni-doped BaFe$_2$As$_2$ \cite{LWHarriger2009}. By comparing the Fig. 5(a) and (b) or Fig. 5(c) and (d), which are plotted to the same color scale, it is clear that the intensities differ little in the superconducting and normal state, consistent with resonance mode being weak. In addition to the magnetic signal, we also observe streaks of spurious scattering of unknown origin [dashed line in Fig. 5(a)]. While the spurious signal overlaps with the magnetic signal at $L=0.5$, the magnetic signal at $L=0$ is reasonably far away. In Fig. 6(a) we show the 1.6 K data subtracted by 12 K data by binning points with $L=0\pm0.2$, which avoids potential contamination from the spurious signal. Similar to Fig. 3(b) and (f), a resonance that is rather broad in energy is seen. The temperature dependence of the resonance mode at $L=0$ is shown in Fig. 6(b), an superconducting-order-parameter-like increase below $T_{\rm c}$ is observed.     

\section{Discussion and Conclusion}

The weak resonance mode in NaFe$_{1-x}$Cu$_x$As is likely a result of the stronger impurity potential of Cu, which is also likely responsible for the lower optimal $T_{\rm c}$ in NaFe$_{1-x}$Cu$_x$As \cite{AFWang2013} compared to NaFe$_{1-x}$Co$_x$As \cite{GTan2016,AFWang2013b} and the near-absence of superconductivity in Ba(Fe$_{1-x}$Cu$_x$)$_2$As$_2$ \cite{NNi2010}. Compared to Ba(Fe$_{1-x}$Cu$_x$)$_2$As$_2$ in which superconductivity is nearly absent, the presence of a superconducting dome in NaFe$_{1-x}$Cu$_x$As may be related to the lower Cu concentration needed to suppress the magnetic order. In a similar vein, lower concentration of Ni in NaFe$_{1-x}$Ni$_x$As was found to cause magnetic order to remain long-range and commensurate approaching optimal superconductivity \cite{CDCao2016}, in contrast to Co- and Ni-doped BaFe$_2$As$_{2}$ in which the magnetic order becomes short-range and incommensurate near optimal-doping \cite{DKPratt2011,HQLuo2012}. The impurity potential and concentration of dopants therefore has significant effects on the physical properties of iron pnictide superconductors. 

The resonance mode in slightly overdoped NaFe$_{0.98}$Cu$_{0.02}$As with $T_{\rm c}=11.8$ K is at $E\approx 5.5$ meV for $L=0.5$, corresponding to $E_{\rm r}\approx 5.5 k_{\rm B}T_{\rm c}$. This is higher than $E_{\rm r}\approx 4.3 k_{\rm B}T_{\rm c}$ suggested for doped BaFe$_2$As$_2$ \cite{DSInosov2011} and $E_{\rm r}\approx 4.5 k_{\rm B}T_{\rm c}$ in slightly overdoped NaFe$_{0.955}$Co$_{0.045}$As \cite{CZhang2013b} but lower than $E_{\rm r}\approx 7.1 k_{\rm B}T_{\rm c}$ in heavily overdoped NaFe$_{0.92}$Co$_{0.08}$As \cite{CZhang2016}. Therefore, due to multi-orbital physics and impurity effects, there appears to be no simple relationship between $E_{\rm r}$ and $T_{\rm c}$ in doped NaFeAs. Specifically, both the concentration of dopants [slightly overdoped and heavily overdoped NaFe$_{1-x}$Co$_x$As] and impurity potential of dopants [slightly overdoped NaFe$_{1-x}$Co$_x$As and NaFe$_{1-x}$Cu$_x$As] seem to increase the ratio $E_{\rm r}/k_{\rm B}T_{\rm c}$. In cuprate superconductors it was found spectral weight of the resonance mode scales linearly with $(E_{\rm c}-E_{\rm r})/E_{\rm c}$ \cite{MKChan2016}, where $E_{\rm c}$ is threshold of the particle-hole continuum. This means as $E_{\rm r}$ moves closer to the particle-hole continuum, spectral weight of the resonance also collapses. A similar effect might also contribute to the weak resonance mode in NaFe$_{0.98}$Cu$_{0.02}$As, as the high ratio of $E_{\rm r}$ and $k_{\rm B}T_{\rm c}$ suggests the resonance is likely closer to the particle-hole continuum compared to NaFe$_{0.955}$Co$_{0.045}$As. 

Similarly, the broader resonance at $L=0$ and 1 compared to $L=0.5$ in NaFe$_{0.98}$Cu$_{0.02}$As [Fig. 3(b), (d) and (f)] may be due to the mode at $L=0$ and 1 being at a higher energy, and experience stronger interactions with particle-hole excitations that broaden the mode. Alternatively, spin-orbit coupling found to be present in many iron pnictide superconductors \cite{Borisenko} and causes energy-splitting of resonance modes polarized along different crystallographic directions \cite{Korshunov13,yusong_BKFA}, also results in broadening of the resonance mode seen in unpolarized neutron scattering experiments. Polarized neutron scattering experiments are needed to distinguish between these scenarios.

The superconducting phase in NaFe$_{1-x}$Cu$_x$As near $x\approx0.02$ can be continuously tuned to the antiferromagnetically ordered insulating phase with significant electronic correlations near $x\approx$0.5 through a region of short-range cation and magnetic order \cite{YSong}, pointing to the possibility of the generic phase diagram of iron pnictides to be anchored around a Mott-insulating state \cite{LdeMedici2014,CYe2015}. The observation of a spin resonance mode in superconducting NaFe$_{1-x}$Cu$_x$As demonstrates that the superconducting state is similar to other iron pnictide superconductors, and therefore electronic correlations should be an integral part of the physics of iron pnictide superconductors. 

\section{Acknowledgments}
The single crystal growth and neutron scattering work at Rice is supported by the
U.S. DOE, BES under contract no. DE-SC0012311 (P.D.). 
A part of the material synthesis work at Rice is supported by the Robert A. Welch Foundation Grant No. C-1839 (P.D.).
Research at BNU and IOP are supported by the National Basic Research Program of China (973 Program, Grants No. 2012CB821401), 
National Natural Science Foundation of China (Grants No. 11374011),
and the Fundamental Research Funds for the Central Universities (Grants No. 2014KJJCB27). Research at Oak Ridge National Laboratory's HFIR was sponsored by the Scientific User Facilities Division, Office of Basic Energy Sciences, U.S. Department of Energy.

\end{document}